\documentclass{article}
\usepackage{fleqn}
\usepackage{epsfig}
\usepackage{amsmath,amssymb}
\begin{document}

\begin{center}
{\bf\Large Statistical system with a fantom scalar interaction in the Gravitation Theory. I. The Microscopic Dynamics.}\\[12pt]
Yu.G. Ignatyev and D.Yu. Ignatyev\\
Kazan Federal University,\\ Kremlyovskaya str., 35,
Kazan 420008, Russia
\end{center}

{\bf keywords}: Early Universe, fantom scalar interaction, relativistic kinetics,
cosmological acceleration, numerical simulation.

\begin{abstract}
Based on the proposed earlier by the Author approach to macroscopic description of scalar interaction, this paper develops the macroscopic model of relativistic plasma with a fantom scalar interaction of elementary particles. In the given model as opposed to previous models a restriction on nonnegativeness of the particles' effective mass is removed.
\end{abstract}

\section{Introduction}
In the recent years there emerged a large number of articles devoted to scalar fields, which are necessary for an explanation of a secondary acceleration of the Universe. It is reasonable to take models with minimal interaction, in which  interaction of a scalar field with an <<ordinary matter>> is implemented by means of a gravitational interaction. Also it is reasonable to take models with non-minimal interaction, in which interaction of a scalar field with an <<ordinary matter>> is implemented by means of some connection. This connection could be realized through the factor depending on the scalar field potential in the Lagrange matter function (potential connection) or on the  factor  which depends on the scalar field's derivatives (kinetic connection). These models are intended to describe to some extent at least qualitative behavior of the Universe at large and small times. However, being essentially highly phenomenological, these models are not able to satisfy aesthetical requirements of theorists and unlikely are long-living. From the other hand, there exist theoretical models based on the gravitation theory's modifications, which are able to describe a kinematics of the cosmological extension.
These are, primarily, $f(R)$ gravitation models of A.Starobinsky \cite{Starobinsky}, and also other modifications of the gravitation theory such as Poincare calibration theories of gravi\-ta\-tion\footnote{see e.g., A.V.Minkevich \cite{Minkevich}.}. By no means negating a possibility of modification of the Einstein theory we nonetheless consider a cosmological model constructed on the fundamental scalar interaction since the scalar field being highly symmetrical one, from one hand can be the foundation of the elementary particle physics and, from the other hand, can realize a concept of physical vacuum. In the article we consider fantom fields, i.e. fields possessing a <<kinetic>> energy.

In contrast to specified nonminimal models of scalar interaction we consider statistical systems of scalar charged particles in which some
sort of particles can directly interact with a scalar field by means of fundamental \textit{scalar charge}. From the other hand, statistical system
possessing  non-zero scalar charge and being itself a source of a scalar field, can efficiently influence on scalar field managing its behavior.
Such a scalar interaction was introduced in the general relativistic theory in 1983 by the Author \cite{Ignatev1,Ignatev2,Ignatev3,Ignatev4}
and a bit later - by G.G.Ivanov \cite{Ivanov}. In particular, in articles %
\cite{Ignatev2,Ignatev3} based on the kinetic theory %
the self-consistent set of equations describing a statistical system of particles with scalar interaction was obtained. In \cite{kuza} there were investigated group properties of equilibrium statistical configurations with scalar interaction and in \cite{Ignat_Popov} it was established a tight connection between the variable mass particle dynamics and the dynamics of scalar charge particle. In recent Author's articles \cite{YuNewScalar1,YuNewScalar2,YuNewScalar3}\footnote{see also monographes \cite{Yubook1,Yubook2}.} the macroscopic theory of statistical systems with scalar interaction was significantly improved and extended to the case of fantom scalar fields\footnote{see also review \cite{YuSTFI}}.

As opposed to the cited above nonminimal models of scalar interaction, the considered strict dynamic model of interaction of a scalar field with elementary particles
is based on the Hamilton mi\-cro\-sco\-pic equations of motion and following statistical averaging procedure. It turns out that
an interaction of a scalar field with particles at that could be introduced by a single way; consequently the ma\-cro\-sco\-pic equations of matter and scalar field
are obtained in a single way by means of standard averaging procedures. Thereby a tight connection between micro and macro levels of a scalar field description is established. Naturally enough that obtained in that way model of scalar interaction should have more complex construction than similar phenomenological models at the same time it should reveal more rich possibilities of behavior\footnote{see e.g. \cite{YuMif}}.

\section{The Canonical Equations Of Motion}
The canonical equations of relativistic particle motion relative to a pair of canonically conjugated dynamic variables $x^{i} $ (coordinates) and $P_{i} $ (generalized momentum) have a form (see e.g. \cite{Ignatev2})\footnote{Here and then the generic unit system is accepted $G=c=\hbar =1$}:

\begin{equation} \label{EQ1}
\frac{dx^{i} }{ds} =\frac{\partial H}{\partial P_{i} } ;\quad \quad \frac{dP_{i} }{ds} =-\frac{\partial H}{\partial x^{i} } ,
\end{equation}
 where $H(x,P)$ is a relativistically invariant Hamilton function. Calculating the full derivative of the dynamic variables $\psi (x^{i} ,P_{k} )$, with an account of  \eqref{EQ1} we obtain:

\begin{equation} \label{EQ2}
\frac{d\psi }{ds} =[H,\psi ],
\end{equation}
where the invariant Poisson brackets are introduced:
\begin{equation} \label{EQ3}
[H,\psi ]=\frac{\partial H}{\partial P_{i} } \frac{\partial \psi }{\partial x^{i} } -\frac{\partial H}{\partial x^{i} } \frac{\partial \psi }{\partial P_{i} } \; .
\end{equation}
As a result of \eqref{EQ3} the Hamilton function is an integral of particle motion:
\begin{equation} \label{EQ4}
\frac{dH}{ds} =[H,H]=0,\Rightarrow H= {\rm Const}.
\end{equation}
The relation \eqref{EQ4} could be named the normalization ratio.
The invariant Hamilton function is determined ambiguously. Indeed, in consequence of  \eqref{EQ3} %
if $H(x,P)$ is a Hamilton function, then any continuously dif\-fe\-ren\-ti\-able function $f(H)$ is also a Hamilton function.
The sole possibility of introduction of the invariant Hamilton function quadratic by generalized particle momentum at presence of only gravitational and scalar fields
is:
\begin{equation} \label{EQ7}
H(x,P)=\frac{1}{2} \left[\psi(x)(P,P)-\varphi(x) \right],
\end{equation}
where $(a,b)$ are here and then a scalar product of 4-dimensional vectors $a$ and $b$, and $\psi(x)$ and $\varphi(x)$ are certain scalar functions.
Choosing a non-zero normalization of the Hamilton function \cite{Ignat_Popov,YuNewScalar1},
\begin{equation}\label{EQ7a}
H(x,P)=\frac{1}{2} \left[\psi(x)(P,P)-\varphi(x) \right]=0.
\end{equation}
we obtain:
\begin{equation}\label{EQ7b}
(P,P)=\frac{\varphi}{\psi},
\end{equation}
and from the first group of the canonical equations of motion(\ref{EQ1}) %
we obtain the relation between the generalized momentum and a particle velocity vector:
\begin{equation} \label{EQ10a}
u^{i} \equiv \frac{dx^{i} }{ds} =\psi P^{i} \Rightarrow P^{i} =\psi^{-1} u^{i} ,
\end{equation}
Substituting the last relation to the normalization ratio (\ref{EQ7b}) and requiring the unit normalization of particle velocity vector
\begin{equation} \label{EQ11}
(u,u)=1.
\end{equation}
we obtain:
\[\psi\varphi=1 \Rightarrow \psi=\varphi^{-1},\]

thereby particle's invariant Hamilton function could be defined by only one scalar function $\varphi(x)$. Taking into account the last relation, let us write down the Hamilton function in the final form:
\begin{equation}\label{EQ7 }
H(x,P)=\frac{1}{2} \left[\varphi^{-1}(x)(P,P)-\varphi(x) \right]=0,
\end{equation}
and from the canonical equations \eqref{EQ1} we obtain the relation between the generalized momentum and particle velocity vector:
\begin{equation}
\label{EQ10} P^i=\varphi \frac{dx^i}{ds}.
\end{equation}

From the definition \eqref{EQ7} it follows that the ge\-ne\-ra\-li\-zed momentum vector is timelike:
\begin{equation}
\label{EQ8} (P,P)=\varphi^2.
\end{equation}
Let us notice useful for future reasoning a relation which is a consequence of  \eqref{EQ3}, \eqref{EQ7} and \eqref{EQ8}:

\begin{equation} \label{EQ9}
[H,P^{k} ]=\nabla ^{k} \varphi \equiv g^{ik} \partial _{i} \varphi.
\end{equation}

\subsection{The Lagrange Equations Of Motion}

From the second group of the canonical equations \eqref{EQ1} we obtain the equations of motion in Lagrange forulation \cite{Yubook1}:

\begin{equation} \label{EQ12}
\frac{d^{2} x^{i} }{ds^{2} } +\Gamma _{jk}^{i} \frac{dx^{j} }{ds} \frac{dx^{k} }{ds} =\partial _{,k} \ln |\varphi|{\rm {\mathcal P}}^{ik} ,
\end{equation}
 where:

\begin{equation} \label{EQ13}
{\rm {\mathcal P}}^{ik} ={\rm {\mathcal P}}^{ki} =g^{ik} -u^{i} u^{k}
\end{equation}
 is a tensor of an orthogonal projection to the direction $u$, such that:

\begin{equation} \label{EQ14}
{\rm {\mathcal P}}^{ik} u_{k} \equiv 0;\quad {\rm {\mathcal P}}^{ik} g_{ik} \equiv 3.
\end{equation}
 From these relations and the Lagrange equations \eqref{EQ12} it follows the strict consequence of velocity and acceleration vectors orthogonality:

\begin{equation} \label{EQ15}
g_{ik} u^{i} \frac{du^{k} }{ds} \equiv 0.
\end{equation}
Let us note that the Lagrange equations of motion (\ref{EQ12}) are invariant relative to a sign of the scalar function $\varphi(x)$:
\begin{equation} \label{EQ16a}
\varphi(x)\rightarrow -\varphi(x).
\end{equation}
The Hamilton function (\ref{EQ7 }) at its zero normalization $H\rightarrow -H$ is also invariant relative to the trans\-for\-ma\-ti\-on (\ref{EQ16a}). Therefore from the relations \eqref{EQ8}, \eqref{EQ10}, and the Lagrange equations \eqref{EQ12} it follows that the quadrate of  $\varphi $ scalar has a meaning of a quadrate of \textit{ the efficient inert mass of particle, $m_{*} $, in a scalar field}:

\begin{equation} \label{EQ16}
\varphi^2 =m_{*}^2 .
\end{equation}
 Let us note that a following action function formally coinciding with the Lagrange function of a rela\-ti\-vis\-tic particle
with a  rest mass $\varphi$ in a gravitational field corresponds to the cited choice of the Hamilton function:
\begin{equation} \label{EQ17}
S=\int m_* ds,
\end{equation}
where an effective mass $m_*$ due to invariance of the Lagrange equations relative to the transformation (\ref{EQ16a}) can be defined as  $m_*=\varphi$, or  $m_*=|\varphi|$.
In this article we resolve the relation (\ref{EQ16}) by next form:
\begin{equation}\label{phi=m}
m_*=\varphi,
\end{equation}
as opposed to how it wa proposed in the previous articles: $m_*=|\varphi|$. Thus at (\ref{phi=m}) an effective mass of a particle in general could be negative.

\subsection{Integrals Of Motion}

 Let us now find conditions of existence of canonical equations' of state \eqref{EQ1} line integral which is con\-nec\-ted to particle's total energy and momentum. For that let us calculate the scalar product's $(\xi ,P)$ full derivative by the canonical parameter. Using the canonical equations of motion \eqref{EQ1}, the normalization ratio \eqref{EQ8}, and the relation of the generalized mo\-men\-tum to the kinematic one \eqref{EQ10}, we find:

\begin{equation} \label{EQ18}
\frac{d(\xi ,P)}{ds} =\frac{1}{m_{*} } P^{i} P^{k} \mathop{L}\limits_{\xi } g_{ik} +\mathop{L}\limits_{\xi } m_{*} ,
\end{equation}
 where $\mathop{L}\limits_{\xi } $ is a Lie derivative by the direction $\xi $ \footnote{see e.g. \cite{Petrov}.}. Assuming further

\begin{equation} \label{EQ19}
\frac{d(\xi ,P)}{ds} =0\Leftrightarrow (\xi ,P)={\kern 1pt} {\rm Const}{\kern 1pt} ,
\end{equation}
 with an account of arbitrariness of the generalized momentum's vector and its normalization ratio we obtain the conditions of the equation fulfillment:

\begin{equation} \label{EQ20}
\mathop{L}\limits_{\xi } g_{ik} =\sigma g_{ik} \Rightarrow \sigma =-\mathop{L}\limits_{\xi } \ln |m_{*} |,
\end{equation}
 where $\sigma$ is a arbitrary scalar. Substituting this result back to the relation \eqref{EQ18}, we find the necessary and sufficient conditions of existence of the canonical equations' line integral (see e.g. \cite{Yubook1}):

\begin{equation} \label{EQ21}
\mathop{L}\limits_{\xi } m_{*} g_{ik} =0.
\end{equation}
 Thus \textit{ to exist the line integral, it is enough and sufficient that conformal corresponding space with metrics $m_{*} g_{ik} $ allows a group of motions with the Killing vector $\xi $}. Let us note that line integrals \eqref{EQ19} have the meaning of total momentum (at spacelike vector $\xi $) or total energy (at timelike vector $\xi $).

\subsection{The Choice Of mass Function}

 There emerges the question of choice of function $m_{*} (\Phi )$. Let us highlight one important circumstance yet not concretizing this function. Let us consider statistical fields $g_{ik} $ and $\Phi $, allowing timelike Killing vector $\xi ^{i} =\delta _{4}^{i} $, when particle's total energy $P_{4} $ is conserved. Further let us consider the reference frame in which  $g_{\alpha 4} =0$, so that coordinate $x^{4} $ coincides with a world time $t$. Then from the relations of connection between the kinematic ve\-lo\-city $u^{i} $ and vector of particle's total momentum $P_{i} $ \eqref{EQ10} if follows:

\begin{equation} \label{EQ22}
P_{4} ds=m_{*} dt,
\end{equation}
 where $P_{4} =E_{0} ={\kern 1pt} {\rm Const}{\kern 1pt} >0$ is a full energy of a charged particle. Therefore if we want to conserve the same orientation of world and proper time (i.e. $dt/ds>0$), it is required to choose such a mass function which always stay nonnegative. In the article we nonetheless remove the requirement of positiveness of an effective mass of particle.\footnote{It should be noted that as early as A.A. Vlasov in his fundamental work \cite{Vlasov} supposed a possibility of discontinuity of a standard connection between these times.}

 Further, from one hand, in absence of a scalar field or, more precisely, in a constant scalar field, mass function should
  change to particle rest mass, $m\ge 0$.
  From the other hand, the Lagrange equations \eqref{EQ12} in case of a weak scalar field and small velocities should change to classical equations of motion in a scalar field:
 \begin{equation}\label{weak_F}
 m\frac{d^2x^\alpha}{dt^2}=-q\partial_\alpha\Phi.
 \end{equation}
 Hence following from the correspondence principle we should have:

\begin{equation} \label{EQ25}
\varphi(0)=m;\quad \lim\limits_{\Phi=0}\left( \frac{\varphi_{,k}}{\varphi}\right)=q\Phi _{,k} ,
\end{equation}
 where $\varphi=\varphi(\Phi)$. %
 Supposing $q\Phi=f$, let us rewrite \eqref{EQ25} in a equivalent form:
 \begin{equation}\label{m_0}
 \varphi(0)=m;\quad \lim\limits_{f=0}\frac{d\varphi}{df}=1,
 \end{equation}
whence \eqref{EQ25} mean that at small values of the scalar potential $\Phi $ function $m_{*} (\Phi )$ should have an ex\-pan\-sion of form:

\begin{equation} \label{EQ26}
m_{*} (\Phi ){\rm \simeq } m+q\Phi+\ldots.
\end{equation}
 To this condition corresponds the linear function
 \begin{equation}\label{m_*}
 m_{*} (\Phi )=m+q\Phi,
\end{equation}
as well as function used in the cited papers:
\begin{equation}\label{|m|}
m_{*} (\Phi )=|m+q\Phi|.
\end{equation}
Further we will choose particle's effective mass function in form (\ref{m_*}). Then according to(\ref{EQ10}) we obtain the relation between
the particle's ge\-ne\-ra\-lized mo\-men\-tum and the kinematic velocity:
\begin{equation}\label{Pu}
P^i=m_*u^i=(m+q\Phi)u^i;
\end{equation}
the normalization ratio for the generalized mo\-men\-tum vector takes form
\begin{equation}\label{norm_m}
(P,P)=m_*^2=(m+q\Phi)^2
\end{equation}
and has an identical form for mass function (\ref{m_*}), as well as for mass function (\ref{|m|}).

To refine the meaning of introduced kinematic and dynamic values let us consider a problem of scalar charged particle's movement in the Minkovsky space in a statistical scalar field, which potential depends only on one coordinate, $x^{1} =x$, and let for the simplicity $m_{*} =m+q\Phi =x$. This we have 3 Killing vectors - one of which is timelike and two are spacelike:
 \def\stackunder#1#2{\mathrel{\mathop{#2}\limits_{#1}}}
$$\stackunder{1}{\xi}^i=\delta _{4}^{i};\; \stackunder{2}{\xi}^i=\delta _{2}^{i} ;\; \stackunder{3}{\xi}^i =\delta _{3}^{i} .$$
Correspondingly to these vectors there exist three linear integrals of motion:

\begin{eqnarray} \label{EQ30}
P_{2} =P_{2}^{0} = {\rm Const} ;\; P_{3} =P_{3}^{0} = {\rm Const};\nonumber\\
 P_{4} =P_{4}^{0} = {\rm Const}.
\end{eqnarray}
 Let for the simplicity  $P_{2} =P_{3} =0$. Then with an account of the normalization ratio \eqref{norm_m}
 we obtain from the canonical equations of motions the non\-tri\-vial one:
\begin{eqnarray} \label{EQ31}
\frac{P^1}{m_*}=u^1\equiv\frac{dx}{ds} =\mp \frac{\sqrt{(P^0_{4})^{2} -x^{2}}}{x }.\\
\label{EQ31a}
u^4\equiv\frac{dt}{ds}\equiv \frac{P_4}{m_*}=\frac{P^0_4}{x}\Rightarrow
\end{eqnarray}
 \begin{equation}\label{Eqx}
 \frac{dx}{dt}=\mp \frac{1}{P^0_4}\sqrt{(P^0_4)^2-x^2},
 \end{equation}
 where signs $\mp$ correspond to a movement to the left or to the right.
 Integrating (\ref{Eqx}) with the initial condition $x(0)=0$,
 we find the solution:
 \begin{equation}
 x=\mp P^0_4\sin(t/P^0_4).
 \end{equation}
 Thus a particle performs harmonic oscillations by clocks of the world time with an amplitude $P^0_4$ and frequency $P^0_4$.
However at that, following (\ref{EQ31a})
 kinematic velocity's components  $u^i$ in moments of time $t=\pi k$ undergo discontinuities of the second kind and change the sign.
  However components of the generalized momentum, $P_i$, %
as well as com\-po\-nents of the kinematic momentum, $p^i$, being the observed variables defined by the standard procedure as
 \begin{equation}\label{p_k}
 p^i=m_*\frac{dx^i}{ds}\equiv P^i,
 \end{equation}
 remain continuous functions and kinematic momentum's component $p^4$ does not change the sign, remaining constant. Nevertheless, the sign of microscopic proper time's relation to the world time is changed. Let us notice, that this outcome is not dependent on the massive function's choice method; the same outcome has been obtained in  \cite{YuNewScalar1}, where positively defined mass function had been being used. Therefore reasonings cited in more recent Author's papers in favor of nonnegativeness of the mass function and formula (\ref{|m|}) on the basis of the Lagrange equations' of motion analysis, are not strict. Apparently, cited example of particle interaction with a scalar field presents the realization of A.Vlasov \cite{Vlasov} concept about the possibility of discontinuity of the relation of microscopic proper time of the particle with the world macroscopic time.

\subsection{Quantum equations}
Let us note that with a use of the standard procedure of quantum equations' obtainment %
from the classical Hamilton function it is required to make a change in the function:\footnote{In %
this place we temporarily depart from universal system of units, in which $\hbar=1$.}
\begin{equation}\label{quant_trans}
P_i\rightarrow i\hbar \frac{\partial}{\partial x^i},
\end{equation}
As a result of covariant generalization, Hamilton operator can be obtained from the Hamilton function:
\begin{equation}\label{Hamiltonian}
\hat{{\rm H}}=-m_*^{-1}(\hbar^2 g^{ik}\nabla_i\nabla_k + m^2_*).
\end{equation}
Thus for a free massive scalar field we could have obtained the wave equations in form of standard Klein-Gordon equations with the only difference that
bosons rest mass should be changed by the effective mass:
\begin{equation}\label{free_bozon}
(\square+m_*^2/\hbar^2)\Psi=0,
\end{equation}
and for free fermions we could have obtained co\-rres\-pon\-ding Dirac equations:
\begin{equation}\label{Dirac_eq}
(\hbar\gamma^i\nabla_i+m_*)\Psi=0,
\end{equation}
where $\gamma$ are spinors.

Let us note that from (\ref{free_bozon}) having substituted $\Psi=\Phi$ and chosen simplest mass function
 $m_*=|q\Phi|$ it right away follows an equation of the free scalar field with
 a cubic nonlinearity:
\begin{equation}\label{Phi3}
\square\Phi+q^2/\hbar^2\Phi^3=0.
\end{equation}
Thus constant of self-action in a scalar field's cubic equation takes quite defined meaning:
$$\lambda=\frac{q^2}{\hbar^2}.$$


\begin{thebibliography}{15}
\bibitem{Starobinsky}
A.A. Starobinsky, Phys. Lett. B {\bf 91 (1)}, 99 (1980).
\bibitem{Minkevich}
 A.V. Minkevich, Gravitation \& Cosmology,  {\bf 12}, 11 (2006).
%
\bibitem{Ignatev1}
Yu.G. Ignat'ev, Russian Physics Journal,
{\bf 25}, No 4, 92 (1982).
\bibitem{Ignatev2}
Yu.G. Ignat'ev, Russian Physics Journal,
{\bf  26}, No 8, 15 (1983).
\bibitem{Ignatev3}
Yu.G. Ignat'ev, Russian Physics Journal,
{\bf  26}, No 8, 19 (1983).
\bibitem{Ignatev4}
Yu.G. Ignat'ev, Russian Physics Journal,
{\bf  26}, No 12, 9 (1983).
\bibitem{Ivanov}
G.G. Ivanov, Russian Physics Journal,
{\bf  26}, No 1, 32 (1983).
%
\bibitem{kuza}
Yu.G. Ignat'ev, R.R. Kuzeev, Ukr. Fiz. J.  {\bf 29}, 1021 (1984).
\bibitem{Ignat_Popov}
Yu.G. Ignatev and A.A. Popov, Actrophysics and Space Science,  {\bf 163}, 153 (1990); arXiv:1101.4303v1 [gr-qc].
%
\bibitem{YuMif}
Yu. Ignat'ev, R. Miftakhov, Gravitation \& Cosmology, {\bf 12}, 179 (2006); arXiv:1011.5774[gr-qc].
%
\bibitem{YuNewScalar1}
Yu. G. Ignat'ev, Russian Physics Journal,
{\bf 55}, No 2, 166 (2012); DOI: 10.1007/s11182-012-9790-9.
%
\bibitem{YuNewScalar2}
Yu. G. Ignat'ev, Russian Physics Journal,
{\bf 55}, No  5, 550 (2012); DOI: 10.1007/s11182-012-9847-9.
%
\bibitem{YuNewScalar3}
Yu. G. Ignat'ev, Russian Physics Journal,
{\bf 55}, No  11, 1345 (2013); arXiv:1307.2509 [gr-qc].
%
%
\bibitem{Yubook1}
Yurii G. Ignatyev, Relativistic
Kinetic Theory of Nonequilibrium Processes in Gravitational
Fields. Kazan, Foliant-Press, -- 2010;  http://rgs.vniims.ru/books/const.pdf.
%
\bibitem{Yubook2}
Yurii G. Ignatyev, The Nonequilibrium Universe: The Kinetics Models of the Cosmological Evolution, Kazan: Kazan University Press, 2013;
http://www.stfi.ru/archive\_rus/2013\_2\_ Ignatiev.pdf
%
\bibitem{YuSTFI}
Yu,G. Ignat'ev, Spase, Time and Foudamental Interections, No 1, 47 (2014) (In Russian).
%
\bibitem{Petrov}
A.Z.Petrov, New Methods in General Theory of Gravitation. Moskow, Nauka, 1966.
%
\bibitem{Vlasov}
A.A. Vlasov, Statistical Distribution Functions. Moskow, Nauka, 1966.
%
\end{thebibliography}
\end{document}